\documentclass[12pt]{iopart}
\usepackage{epsfig}
\usepackage{savesym}
\expandafter\let\csname equation*\endcsname\relax
\expandafter\let\csname endequation*\endcsname\relax
\usepackage{amsmath}
\usepackage[latin1]{inputenc}
\usepackage{amsbsy}
\usepackage{color}
\usepackage{epsfig}
\usepackage{graphicx}
\usepackage{amsmath}
\usepackage{amssymb}
\usepackage{bbm}
\usepackage{amsbsy}

\newcommand{\arctanh}{\operatorname{arctanh}}
\usepackage{iopams}

\begin{document}

\title[M. Loewe, F. Marquez and R. Zamora]{The cylindrical $\delta-$potential and the Dirac equation}

\author{M. Loewe, F. Marquez and R. Zamora}

\address{Facultad de F\'{\i}sica, Pontificia Universidad Cat\'{o}lica, Casilla 306, Santiago 22,
Chile}
\ead{mloewe@fis.puc.cl, cfmarque@uc.cl and rrzamora@uc.cl}

\begin{abstract}

In this article we discuss the Dirac equation in the presence of an
attractive cylindrical $\delta$-shell potential
$V(\rho)=-a\delta(\rho-\rho_0) ,$ where $\rho$ is the radial
coordinate and $a>0.$  We present a detailed discussion on the
boundary conditions the wave function has to satisfy when crossing
the support of the potential, proceeding then to explore the
dependence of the ground state on the parameter $a$, analyzing the
occurrence of supercritical effects. We also apply the
Foldy-Wouthuysen transformation, discussing the non-relativistic
limit of this problem.

\end{abstract}
\maketitle

\section{Solution of the Dirac equation for a cylindrical $\delta$-shell}

The Dirac equation opened certainly a new world to physics, with
marvelous and surprising consequences as, for example, the existence
of anti-matter, the fact that $g=2$ for the electron giromagnetig
moment, etc. \cite{Dirac1}-\cite{Thaller}. However, perhaps one of
the most interesting facts associated with the Dirac equation is the
idea of the existence of a populated vacuum, the Dirac sea, i.e. a
vacuum state having a non trivial structure. This concept developed
later into a  cornerstone of our present understanding of modern
quantum field theory.  Remnants of this exciting idea are, for
example, the vacuum condensates in Quantum Chromodynamics, the
thermal vacuum distribution, when finite temperature effects are
taking into account in heavy ion-collisions, and, in general, the
idea of non-zero vacuum expectation values. Supercritical effects,
i.e. an instability that appears when the ground state starts to
dive into the Dirac sea, inducing then positron emission, is a
prediction of this scenario, which has nevertheless not been
experimentally confirmed yet \cite{greiner}.

Singular $\delta $-type potentials have been considered in quantum
mechanics already from the beginning. In general, singular $\delta
$-type potentials can be considered as toy models which allow us to
get a physical insight, being, at the same time, more easy to deal
with in comparison with more realistic extended potentials.  An
example is the well known Dirac-comb or Kronig-Penney model
\cite{KP} in non relativistic quantum mechanics, which provided us
with an understanding and intuition about the emergence of band
structures in solid state physics. Another situation where this type
of interactions have been used, is in nuclear physics as a model
for residual interactions between nucleons inside an incomplete nuclear
shell, \cite{nuclear}, providing a correct qualitative picture to
this difficult problem.

In the frame of the Dirac equation, singular $\delta $-type
interactions have also been considered many times in the literature
\cite{BCL}, \cite{Danilo}. Certainly this problem is also attractive
from the perspective of mathematical physics. A rigorous
construction of self adjoint extensions for the Dirac operator,
allowing the handling of matching conditions at the support of the
$\delta $-potential for the spherically symmetric case were
considered in \cite{exner}. These conditions are essential for
determining the existence of bound states. The theorem by Svendsen
\cite{svendsen} is also a rigorous result which tells us that for
the spherically symmetric singular potential $V(\rho) = - a
\delta(\rho - \rho _{0})$, supercritical effects will be absent in
the limit where $\rho _0 \rightarrow 0$. This fact is a particular
case of the general result of this theorem, namely that contact
interactions for the Dirac equation can be constructed only if the
supporting manifold has a codimension of dimension one, which is the
case in our problem. See also  \cite{nogami} in this context. The
cylindrical attractive $V(\rho) = - a \;\delta (\rho-\rho_{0})$
potential, the main subject of this article, as we will see, presents
many interesting features, being worthwhile to go into a detailed
discussion of its properties. This case has been discussed in the
literature but in the context of the $( 2+1 )$ Dirac equation, where
we have three two by two Dirac matrices \cite{shi-hai}. The
treatment here is different, and relies on an appropriate unitary
transform and the use of the chiral representation for the Dirac
matrices.

\medskip
If we think about the interaction of a neutral Dirac particle, which
carries a magnetic moment $\mu $, for example a neutron,  with an
external magnetic field $\vec{B}$, $H_{I} = - \vec{\mu }\cdot
\vec{B}$, we will have a point-like interaction of a $\delta $-type
if the external magnetic field corresponds to an Aharonov-Bohm
magnetic vortex. The situation we propose to discuss in this article
corresponds physically to an array of Aharonov-Bohm vortices,
distributed along a circle of radius $\rho_{0}$, which may interact
with the magnetic moment of a neutral Dirac particle. Several
experiments have been done on the behavior of electrons under the
influence of external Aharonov-Bohm magnetic fields \cite{tonomura},
and such an array could in principle be constructed.

\noindent
 Obviously, we are compelled to work in cylindrical
coordinates. Since a $\delta$-type potential, which has support in a
domain of zero measure, divides the space into two regions, we will
start by considering first the free particle case. Then, in a second
step we will establish the connection between the wave functions
when crossing the support of the potential.


\subsection{Dirac free particle in cylindrical coordinates}

The Dirac equation in curvilinear coordinates, in general, includes a non trivial spin connection in the covariant derivative \cite{Nakahara}. When considering the free particle case in cylindrical coordinates,
it turns out that the relevant Dirac matrices are coordinate
dependent because here we have

\begin{equation}
\sigma_\rho=\begin{pmatrix} 0 & e^{-i\theta} \\ e^{i\theta} & 0 \end{pmatrix} \mbox{\hspace{10pt};\hspace{10pt}}\sigma_\theta=\begin{pmatrix}0 & -ie^{-i\theta} \\ ie^{i\theta} & 0 \end{pmatrix}.
\end{equation}

This is something we would like to avoid when solving the free Dirac
equation $(\gamma \cdot p -m) \psi = 0.$ To deal with this, let us consider the unitary transformation
\begin{equation}\hat{S}=\frac{1}{\sqrt{\rho}}e^{\frac{\theta}{2}\gamma^1\gamma^2},\end{equation}
\noindent introduced in \cite{Villalba}. Applying this transformation to the Dirac equation and defining $\psi^{\prime}\equiv \hat{S}\psi$, we get
\begin{equation}
(\gamma^0\partial_t+\gamma^1\partial_\rho+\frac{\gamma^2}{\rho}\partial_\theta+\gamma^3\partial_z+im)\psi'=0.
\end{equation}
Notice that in our case, due to the form of $\hat{S}$, the spin connection term is trivial. At this point we can see the effect of the unitary transformation
$\hat{S}$. It has rotated our equation so that the $\gamma$ matrices
involved are the cartesian ones, eliminating their dependence on the
coordinates.\\
We now introduce the Ansatz $\psi'=\gamma^1\gamma^2\phi$, and rewrite equation (3) as
\begin{equation}\left(\hat{H}_1+\hat{H}_2\right)\phi=0,\end{equation}
where
\begin{eqnarray}
\hat{H_1}&=&\left(\gamma^0\partial_t+\gamma^3\partial_z\right)\gamma^1\gamma^2\\
\hat{H_2}&=&\left(\gamma^1\partial_\rho+\frac{\gamma_2}{\rho}\partial_\theta+im\right)\gamma^1\gamma^2.
\end{eqnarray}
It is easy to see that these operators commute. This means that if each one of them satifies an eigenvalue problem, they have a common basis, i.e.
\begin{eqnarray}
\hat{H}_1\phi=\lambda_1\phi\\
\hat{H}_2\phi=\lambda_2\phi.
\end{eqnarray}
Putting this into equation (4), it is trivial to note that $\lambda_1=-\lambda_2\equiv\lambda$. In this way we can write two equations for the 4-component bispinor $\phi$
\begin{eqnarray} (-\gamma^2\partial_\rho+\frac{\gamma^1}{\rho}\partial_\theta+im\gamma^1\gamma^2+\lambda)\phi=0 \\
(\gamma^0\gamma^1\gamma^2\partial_t+\gamma^3\gamma^1\gamma^2\partial_z-\lambda)\phi=0.
\end{eqnarray}
It is natural then to propose
\begin{equation} \phi=e^{-iEt}e^{ik_zz}e^{ik_\theta \theta} \begin{pmatrix} \varepsilon \\ \eta \end{pmatrix} ,
\end{equation}
\noindent where $\varepsilon$ and $\eta$ are both 2-component spinors. Note that, because of the symmetry of the problem we would expect to have a plane wave solution in the $z$ coordinate. This ansatz is quite general, precisely, due to the strong symmetry of our problem. Since we are imposing no restriction on the radial coordinate $\rho$, this ansatz should represent a general and complete solution of the problem. For our purpose it is
convenient to use the Chiral representation for the Dirac matrices
 \cite{Chiral} given by
\begin{equation}\gamma^0=\begin{pmatrix} 0 & I \\ I & 0 \end{pmatrix}
\hspace{1cm},\hspace{1cm}\gamma^i=\begin{pmatrix} 0 & \sigma_i \\
-\sigma_i & 0 \end{pmatrix} .\end{equation}

\noindent Substituting then our Ansatz in the previous equations we
get
\begin{eqnarray}
iE\sigma^1\sigma^2\eta+ik_z\sigma^3\sigma^1\sigma^2\eta-\lambda\varepsilon=0 \nonumber\\
iE\sigma^1\sigma^2\varepsilon-ik_z\sigma^3\sigma^1\sigma^2\varepsilon-\lambda\eta=0.
\end{eqnarray}

\noindent It's important to emphasize that even though we have a
rotated spinor and we are working in the Chiral basis, our energy
levels remain unchanged since unitary transformations  do not affect
the energy spectrum. We define

\begin{equation} \varepsilon=\begin{pmatrix} \varepsilon_1 \\ \varepsilon_2 \end{pmatrix} \mbox{\hspace{10pt} ; \hspace{10pt}} \eta=\begin{pmatrix} \eta_1 \\ \eta_2 \end{pmatrix},\end{equation}
where now, $\varepsilon_{1,2}$ and $\eta_{1,2}$ are scalar complex functions (i.e. each of them represents one of the components of the bispinor $\phi$). Replacing this on the previous equation, we can obtain the following relations
\begin{eqnarray} \lambda^2&=&E^2-k_z^2 \nonumber\\
\varepsilon_1&=&-\frac{E+k_z}{\lambda}\eta_1 \\
\varepsilon_2&=&\frac{E-k_z}{\lambda}\eta_2.\nonumber
\end{eqnarray}
These relations are quite simple. Their derivation would have been
more involved if the normal Dirac representation for the Dirac
matrices would have been used. From equations (9) and (10), using
the previous relations, we get

\begin{eqnarray}
\eta'_2=-\frac{k_\theta}{\rho}\eta_2+i\frac{(m+\lambda)(E+k_z)}{\lambda}\eta_1 \\
\eta'_1=\frac{k_\theta}{\rho}\eta_1-i\frac{(m-\lambda)(E-k_z)}{\lambda}\eta_2,
\end{eqnarray}
where a prime denotes derivative with respect to $\rho$.
\\
Without loss of generality, we can take $k_z=0$ decoupling this set
of equations. We define $r\equiv\xi\rho$, where $\xi^2\equiv m^2-\lambda^2$. It is important to notice that $\xi^2$ is positive
definite, since we are looking for bound states, i.e. $-m<E<m$. In terms of this new variable we can write

\begin{equation} r^2\eta''_2-(r^2+k_\theta(k_\theta+1))\eta_2=0, \end{equation}

\medskip
\noindent The previous equation is the well-known Bessel equation of
second kind. In this way,  $\eta_1$ and $\eta_2$
are given by
\begin{eqnarray}
\eta_2&=&\rho^{\frac{1}{2}}(A I_{k_\theta+1/2}(\xi\rho)+B K_{k_\theta+1/2}(\xi\rho)) \nonumber\\
\eta_1&=&i\frac{\lambda}{E}\left(\frac{m-\lambda}{m+\lambda}\right)^\frac{1}{2}\rho^\frac{1}{2}(-A I_{k_\theta-1/2}(\xi\rho)+B K_{k_\theta-1/2}(\xi\rho))\\
&\equiv &i\chi\rho^{\frac{1}{2}}(-A I_{k_\theta-1/2}(\xi\rho)+B
K_{k_\theta-1/2}(\xi\rho)).
\end{eqnarray}
Once we have found the solution for the Dirac free particle in
cylindrical coordinates, we can return to our original problem, the
attractive $\delta$-shell potential. The boundary conditions when
crossing the support of the $\delta$ potential, to be discussed in
the next section,   will provide the energy spectrum.


\subsection{Boundary conditions for the Dirac equation in the presence of a cylindrical $\delta$-shell}

By applying the transformation $\hat{S}$ to the Dirac equation, $(\gamma\cdot p-m+\gamma^0 a\delta(\rho-\rho_0))\psi=0$, we get
\begin{equation} (\gamma^0\partial_t+\gamma^1\partial_\rho+\frac{\gamma^2}{\rho}\partial_\theta+\gamma^3\partial_z+im-ia\gamma^0\delta(\rho-\rho_0))\psi'=0 .\end{equation}
Following exactly the same procedure employed previously,  we now get
\begin{eqnarray} 
&\Leftrightarrow& -i\partial_\rho\eta_2-i\frac{k_\theta}{\rho}\eta_2-im\varepsilon_1-a\delta(\rho-\rho_0)\eta_1+\lambda\varepsilon_1=0 \\
&{}&
i\partial_\rho\eta_1-i\frac{k_\theta}{\rho}\eta_1+im\varepsilon_2+a\delta(\rho-\rho_0)\eta_2+\lambda\varepsilon_2=0.
\end{eqnarray}
From here we get
\begin{multline} \frac{1}{\eta_2^2/\eta_1^2-1}\frac{d}{d\rho}\left(\frac{\eta_2}{\eta_1}\right)+ia\delta(\rho-\rho_0)\\=
\frac{1}{\eta_2^2-\eta_1^2}\left(i\frac{2k_\theta}{\rho}\eta_1\eta_2+
im(\varepsilon_1\eta_1-\varepsilon_2\eta_2)-\lambda(\varepsilon_1\eta_1+\varepsilon_2\eta_2)\right)
.\end{multline}
\noindent Integrating between the limits
$[\rho_0-\epsilon, \rho_0+\epsilon]$, we have
\begin{equation} - \arctanh\left(\frac{\eta_2}{\eta_1}\right)
|_{\rho_0-\epsilon}^{\rho_0+\epsilon} + ia =0.\end{equation}

To keep going, we divide the space in two regions:
\\
Region I: $\rho<\rho_0$
\begin{align}
\eta_1^I=-i\chi\rho^\frac{1}{2} A I_{k_\theta-1/2} \\
\eta_2^I=\rho^\frac{1}{2} A I_{k_\theta-1/2}.
\end{align}
\\
Region II: $\rho>\rho_0$
\begin{align}
\eta_1^{II}=-i\chi\rho^\frac{1}{2} B K_{k_\theta-1/2} \\
\eta_2^{II}=\rho^\frac{1}{2} B K_{k_\theta-1/2}.
\end{align}
The functions in regions I and II are different because the Dirac
four-spinor has to be normalized. In this way we get the relation
\begin{equation}i\frac{\frac{\eta_2^{II}(\rho_0+\epsilon)}{\eta_1^{II}(\rho_0+\epsilon)}-\frac{\eta_2^{I}(\rho_0-\epsilon)}{\eta_1^{I}(\rho_0-\epsilon)}}{1-\left(\frac{\eta_2^{II}(\rho_0+\epsilon)}{\eta_1^{II}(\rho_0+\epsilon)}\right)\left(\frac{\eta_2^{I}(\rho_0-\epsilon)}{\eta_1^{I}(\rho_0-\epsilon)}\right)}=-\tan{a}.\end{equation}
We are interested in the ground state energy ($k_\theta=0$). Using
the explicit expressions of the relevant Bessel functions \cite{Abramowitz}, and taking the limit $\epsilon\rightarrow 0$, we get the
transcendental equation that determines the behavior of the ground state
\begin{equation}\label{gs} \frac{-\sqrt{\frac{1-\varepsilon}{1+\varepsilon}}+\tanh{(s_0)}}{1-\frac{1+
\varepsilon}{1-\varepsilon}\tanh{(s_0)}}=-\tan{a},\end{equation}
where we have defined the following dimensionless variables

\begin{eqnarray}
\varepsilon&=&\frac{E}{m}\nonumber\\
\varphi&=&m\rho_0\\
s_0&=&\varphi\sqrt{1-\varepsilon}.\nonumber
\end{eqnarray}

\medskip
\noindent  For the  derivation of this equation it was important to
fix the value of the constant $\lambda$. Its value can either be
$\pm|E|$. In order to get rid of this ambiguity we notice that when
$a=0$ (no potential), the energy of the electron should correspond
to the free particle energy. To accomplish this $\lambda$ has to be
equal to $-|E|$. Finally, we can remove any ambiguity by demanding a
continuous energy spectrum when $E=0$. This demands $\lambda=-E$.

\medskip
\noindent The evolution of the ground state energy as  function of
the different parameters involved is shown in Fig.1. Numerically it
seems that there will be no critical effects, i.e
$\frac{d\varepsilon }{da} \rightarrow 0$ when $\varepsilon
\rightarrow -1$. In fact it is easy to show this property
analytically. If we consider eq. (31) and define $1+ \varepsilon =
\tilde{\varepsilon}$, then we need to consider the limit when
$\tilde{\varepsilon}\rightarrow 0$. It is easy to find for the
leading behavior of the derivative

$$\frac{d\tilde{\varepsilon}}{da} = \frac{\tilde{\varepsilon}}{\cos
^{2}a(\tanh (\sqrt{2}m\rho _{0}) + \tan a)} \rightarrow 0$$

\noindent when $\tilde{\varepsilon} \rightarrow 0$, i.e.
supercritical effects in this case are absent, in contrast to what
occurs in the case of the spherical delta potential discussed in
\cite{BCL}.

\begin{figure}[!h]
\centering
\includegraphics[scale=0.5]{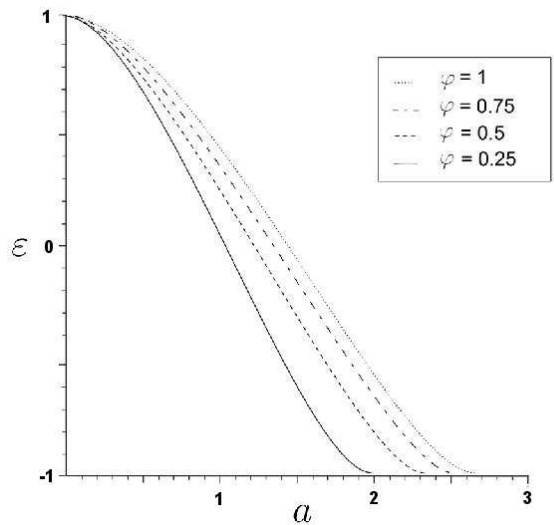}
\caption{The energy of the ground state for a $\delta$-shell potential
as a function of the coupling a.}
\end{figure}

\section{Foldy-Wouthuysen transformation for the $\delta$-shell}

We would like now to discuss the Foldy-Wouthuysen\cite{Foldy} transformation in
the case of our $\delta$-shell problem. In this section we will
obtain some transcendental equations for the energy of the ground
state up to order $1/m$ and $1/m^2$ separately. Later, we will
compare this energy with the energy provided by the Schrödinger
equation. In addition, we will show how the spin-orbit coupling term
appears from the Foldy-Wouthuysen transformation up to order
$\frac{1}{m^2}$.

\subsection{Foldy-Wouthuysen transformation for the Hamiltonian}

We have the hamiltonian
\begin{equation} \hat{H}=\vec{\alpha}\cdot\vec{p}+\beta m-a\delta(\rho-\rho_0) .\end{equation}
We apply a first Foldy-Wouthuysen trasformation $F=e^{iS}$, with $S=-i\beta\frac{\vec{\alpha}\cdot\vec{p}}{2m}$, and then a second one with $S^\prime=-i\frac{\beta\vec{\alpha}\cdot\vec{p}}{2m}(\xi_0-\frac{a}{2m}a\delta(\rho-\rho_0))$
where, similarly to the previous one, all odd operators in the hamiltonian have been included inside
$S^\prime$. We also have defined $\xi_0=\frac{\vec{p}\hspace{0.05cm}^2}{3m^2}$. In this manner we arrive to
\begin{multline} H^{\prime\prime}=\beta\left(m+\frac{\vec{p}\hspace{0.05cm}^2}{2m}\right)-a\delta(\rho-\rho_0)\\-\frac{a}{8m^2}(2(\vec{\alpha}\cdot\vec{p}~\delta(\rho-\rho_0))\alpha_\theta p_\theta-\vec{\alpha}\cdot\vec{p}~(\vec{\alpha}\cdot\vec{p}~\delta(\rho-\rho_0))) \end{multline}
We can see that $H^{\prime\prime}$ has no odd operators, and so it
should be the non relativistic limit of the Hamiltonian up to order
$1/m^2$. It is not difficult to see that, in spite of the presence
of two $\alpha $ matrices,  the last two terms in equation (34)
correspond to even operators.

\subsection{Solving the free particle in the non relativistic limit}

Now we solve the free particle using the hamiltonian $H^{\prime\prime}$ without the Dirac delta potential. We start from
\begin{equation} i\frac{\partial\psi}{\partial t}=\beta\left(m+\frac{\vec{p}\hspace{0.05cm}^2}{2m}\right)\psi, \end{equation}
and suggest the ansatz $\psi=e^{-iEt}\begin{pmatrix}u\\v\end{pmatrix}$, with $u=e^{ik_\theta}u(\rho)$, to get
\begin{equation}r^2u^{\prime\prime}+ru^\prime-k_\theta^2u-\kappa^2u=0,\end{equation}
where $\kappa^2\equiv-2m(E-m)$ and $r\equiv\kappa\rho$.

It is important to note that, since we are looking for bound states,
i.e. $-m<E<m$, $\kappa^2$ is positive definite.

Equation (36) is the Bessel equation, and its
solutions are well known. We can follow a completely similar process
for $v$ to get the solutions
\begin{eqnarray}
u=A I_{k_\theta}(\kappa\rho)+B K_{k_\theta}(\kappa\rho)\\
v=C I_{k_\theta}(\sqrt{2m(E+m)}\rho)+D
K_{k_\theta}(\sqrt{2m(E+m)}\rho).
\end{eqnarray}
Now that we have the solutions, we can apply the boundary
conditions, associated to the $\delta$ potential.

\subsection{Boundary conditions for the non relativistic limit of the $\delta$-shell}

We have the Dirac equation, up to order $1/m$
\begin{equation} i\frac{\partial\psi}{\partial t}=\left(\beta m+\beta\frac{\vec{p}\hspace{0.05cm}^2}{2m}-a\delta(\rho-\rho_0)\right)\psi .\end{equation}
For the upper component, we get
\begin{equation} -\kappa^2u=u^{\prime\prime}+\frac{1}{\rho}u^\prime-\frac{k_\theta}{\rho^2}u+2ma\delta(\rho-\rho_0)u. \end{equation}
We are interested in the ground state ($k_\theta=0$), so the
previous equation becomes
\begin{equation} -\kappa^2u=u^{\prime\prime}+\frac{1}{\rho}u^\prime+2ma\delta(\rho-\rho_0)u .\end{equation}
We integrate between [$\rho_0-\epsilon$, $\rho_0+\epsilon$]. We
should keep in mind that, since the equation is a second order
differential equation, unlike the usual Dirac equation, our wave
function is continuous in $\rho_0$:
\begin{equation}\left. u^\prime\right|_{\rho_0-\epsilon}^{\rho_0+\epsilon}+2mau(\rho_0)=0.\end{equation}
Once again, we divide the space in two regions.

Region I ($\rho<\rho_0$):
\begin{eqnarray}
u_I&=&A I_0(\kappa\rho)\\
v_I&=&C I_0(\sqrt{2m(E+m)}\rho).
\end{eqnarray}

Region II ($\rho>\rho_0$):
\begin{eqnarray}
u_{II}&=&B K_0(\kappa\rho)\\
v_{II}&=&C K_0(\sqrt{2m(E+m)}\rho).
\end{eqnarray}
These functions are different in each region in order to assure
normalization of the Dirac spinor.

With this, we can write, taking the limit $\epsilon\rightarrow0$
\begin{equation}u^\prime_{II}-u^\prime_{I}+2mau(\rho_0)=0.\end{equation}
Imposing continuity of $u$ in $\rho_0$ we can get the relation
\begin{equation}
A=B\frac{K_0(\kappa\rho_0)}{I_0(\kappa\rho_0)}.
\end{equation}
Using this result and defining the dimensionless variables
\begin{eqnarray}
\varepsilon&=&\frac{E}{m}\nonumber\\
\varphi&=&\rho_0m.
\end{eqnarray}
we can write
\begin{multline}\sqrt{2(1-\varepsilon)}K_1(\kappa\rho_0)I_0(\kappa\rho_0)\\+\sqrt{2(1-\varepsilon)}K_0(\kappa\rho_0)I_1(\kappa\rho_0)-2aK_0(\kappa\rho_0)I_0(\kappa\rho_0)=0, \end{multline}
which is a transcendental equation for the energy of the ground
state, up to order $1/m$.
\\
We now repeat the previous procedure up to order $1/m^2$. We have
the Dirac equation
\begin{multline} i\frac{\partial\psi}{\partial t}=\left(\beta\left(m+\frac{\vec{p}\hspace{0.05cm}^2}{2m}\right)-a\delta(\rho-\rho_0)\right.\\\left.-\frac{a}{8m^2}(2(\vec{\alpha}\cdot\vec{p}~\delta(\rho-\rho_0))\alpha_\theta p_\theta-\vec{\alpha}\cdot\vec{p}~(\vec{\alpha}\cdot\vec{p}~\delta(\rho-\rho_0)))\right)\psi, \end{multline}
where
\begin{equation}2(\vec{\alpha}\cdot\vec{p}~\delta(\rho-\rho_0))\alpha_\theta p_\theta=-2\delta^\prime(\rho-\rho_0)\alpha_\rho\alpha_\theta \nabla_\theta=-2\delta^\prime(\rho-\rho_0)\alpha_z \nabla_\theta.\end{equation}
If we focus in the upper component, this term becomes:
\begin{equation}-2\delta^\prime(\rho-\rho_0)\sigma_z\nabla_\theta=-4\delta^\prime(\rho-\rho_0)\hat{L}_z\hat{S}_z.\end{equation}
We recognize the spin-orbit coupling, which is one of the
relativistic corrections provided by the Foldy-Wouthuysen
transformation. However this term has no influence on  the ground
state, where $L_z=k_\theta=0$. We now proceed to solve equation
(51). Using the same ansatz as before for the upper component, we
get
\begin{multline}Eu=mu-\frac{u^{\prime\prime}}{2m}-\frac{u^\prime}{2m\rho}+\frac{k_\theta^2}{2m\rho^2}u-a\delta(\rho-\rho_0)u\\-\frac{a}{4m^2}\frac{k_\theta\sigma_z}{\rho}\delta^\prime(\rho-\rho_0)u-\frac{a}{8m^2}\left(\delta^{\prime\prime}(\rho-\rho_0)+\frac{\delta^\prime(\rho-\rho_0)}{\rho}\right)u.\end{multline}
We solve for the ground state ($k_\theta=0$) and the previous
equation becomes
\begin{multline}\frac{\rho u^{\prime\prime}}{2m}+\frac{u^\prime}{2m}+\rho a\delta(\rho-\rho_0)u\\+\frac{a\rho}{8m^2}\left(\delta^{\prime\prime}(\rho-\rho_0)+\frac{\delta^\prime(\rho-\rho_0)}{\rho}\right)u+(E-m)u=0 .\end{multline}
Integrating between [$\rho_0-\epsilon$, $\rho_0+\epsilon$] and
taking the limit $\epsilon\rightarrow0$, we get
\begin{equation} \frac{1}{2m}(u^\prime_{II}(\rho_0)-u^\prime_I(\rho_0))+au(\rho_0)+\frac{a}{8m^2}u^{\prime\prime}_{II}(\rho_0)+\frac{a}{8m^2}u^\prime_{II}(\rho_0)=0 .\end{equation}

After inserting solutions (43) and (45), equation (56) becomes
\begin{multline}-\sqrt{2(1-\varepsilon)}I_0(\kappa\rho_0)K_1(\kappa\rho_0)+\left(a+\frac{a}{4}(1-\varepsilon)\right)I_0(\kappa\rho_0)K_0(\kappa\rho_0)\\-\sqrt{\frac{1-\varepsilon}{2}}I_1(\kappa\rho_0)K_0(\kappa\rho_0)=0 .\end{multline}
This is yet another transcendental equation for the energy of the
ground state, up to order $1/m^2$.

From equations (57) and (50), we can see how the ground state energy depends on the coupling constant $a$.

\begin{figure}[!h]
\centering
\includegraphics[scale=0.8]{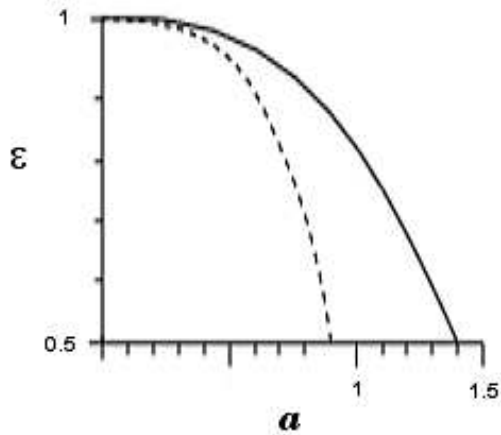}
\caption{Behaviour of the ground state as a function of a. The solid
line is the solution to Schrödinger's equation and the dashed line
is the second order Foldy-Wouthuysen approximation, both for
$\varphi=1$.}
\end{figure}

Although we have not mentioned it here, the solid line stands for both the $1/m$ order of the non relativistic limit and the Schrödinger equation, (after the rest energy has been considered), so there is no relativistic correction, other than the rest energy, to the Schrödinger equation at first order in the Foldy-Wouthuysen transformation. However, up to order $1/m^2$, the first relativistic corrections begin to appear. As we discussed, the spin-orbit coupling can be explicitly written in the equation. Other corrections must be present to, since the energy of the groundstate decreases for the same coupling value, with respect to the non relativisitic solution.

\section{Conclusions}

We have discussed the solutions of the Dirac equation for an
atractive cylindrical delta potential. It turns out that there are
no supercritical effects, in contrast to what happens in the
spherical symmetric case. We discussed also the non relativistic
limit of these problems according to the Foldy-Wouthuysen
approximation. No relativistic corrections are found up to order $1/m$, however, up to order $1/m^2$ relativistic corrections are present, such as the spin-orbit coupling.

\section{Acknowledgements}

The authors acknowledge support from FONDECYT under grant
Nr.1095217. M.L. acknowledges also support from the Proyecto Anillos
ACT119. F.M. and R.Z. acknowledge support from CONICYT under grants 21110577 and 21110295. The authors would like to thank M. Bañados for helpful discussions about the spin connection.

\end{document}